# An Export Architecture for a Multimedia Authoring Environment

Jan Mikáč
INRIA Rhône-Alpes
655 avenue de l'Europe
38334 Saint Ismier, France
+33 076 61 54 38

Jan.Mikac@inria.fr

Cécile Roisin
UPMF & INRIA Rhône-Alpes
655 avenue de l'Europe
38334 Saint Ismier, France
+33 076 61 53 60

Cecile.Roisin@inria.fr

Bao Le Duc
Université Pierre et Marie Curie
15, Rue Ecole de Médecine
75006 Paris, France
+33 637 31 04 60

leducbao@gmail.com

## ABSTRACT
In this paper, we propose an export architecture that provides a clear separation of multimedia authoring services from publication services. We illustrate this architecture with the LimSee3 authoring tool and several standard publication formats: Timesheets, SMIL, and XHTML.

## Categories and Subject Descriptors
I.7 [Document and Text Processing]: Document Preparation — Hypertext/hypermedia, Multi/mixed media, Standards

## General Terms
Documentation, Experimentation, Standardization.

## Keywords
Export, multimedia document, publishing format, SMIL, Timesheets

## 1. INTRODUCTION
The rise of rich web applications in recent years brings many challenges to researchers regarding multimedia authoring, publishing formats and multimedia document rendering. A multimedia authoring system dedicated to end-users aims at facilitating multimedia documents creation. It is worth noting that multimedia authoring is a complex process which demands users to specify document content from different sources, together with their spatial layout, their synchronization (temporal layout) and their behavior on user interactions [2]. A number of available tools support multimedia authoring, including commercial software such as Adobe Flash Creative Suite 3, SwiSH, PowerPoint and open source tools such as GRiNS [5] for SMIL, or Sprout for Flash. These tools are usually tightly coupled with publishing formats. Publishing formats allow one to express multimedia documents under executable formats, possibly taking into account player/system configuration. They can follow open standards such as HTML (with Javascript), SVG, SMIL, XMT or proprietary formats (usually binary formats) such as Flash (with ActionScript) or PowerPoint. The rapid spread of rich web applications that now cover various domains (leisure, education, trading, advertising, or simply individual communication) together with the quasi-permanent emergence of new multimedia technologies pave the way to separate authoring services from publication formats.

In this paper we propose an architecture that enables such a separation. Section 2 identifies the needs of an exportation service, section 3 briefly presents the authoring context in which this work was done, section 4 describes the exportation architecture and section 5 illustrates the benefits gained through two experiences of publication.

## 2. EXPORTATION NEEDS
The authoring services that are provided by authoring tools have to be completed by a set of publishing services (also called exportation services) to cope with the different publication formats in which the users want their multimedia documents to be accessed.

This approach brings up two main benefits :

- Better authoring services because they can better target user needs.
- Independence of the authoring tool from the publication formats.

This last benefit is very important because multimedia document formats are continuously evolving, even those that are defined by standard organizations such as W3C (SMIL, SVG, HTML) or ISO/IEC (MPEG4). This independence ensures therefore a more stable perspective for the document created with the authoring tool.

It has also another interesting advantage for users: the separation between authoring model and publication format allows them to choose the output process adapted to each context where the multimedia information has to be delivered. The choice can also be driven by the kind of multimedia documents being produced. For instance, lightly-synchronized documents can be exported to XHTML+JavaScript in order to provide a wide access (only a web browser is required) while a SMIL-based solution is required when more complex scheduling is necessary.

## 3. THE LIMSEE3 AUTHORING TOOL
LimSee3 [7] is a *generic tool* (or platform) for editing multimedia documents and as such it provides several general authoring mechanisms. The underlying document model [3] is designed to capture author's view of a multimedia document independently of a particular presentation format or player.

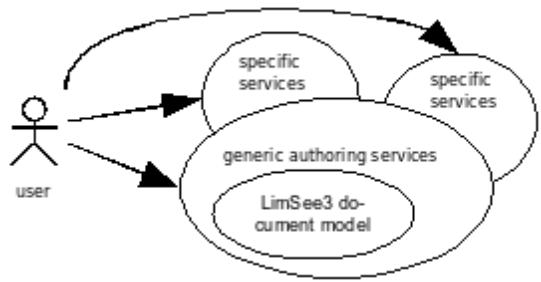

**Figure 1. LimSee3 generic/specific architecture**

On top of the generic platform, LimSee3 proposes some *domain-* or *application-specific* enhancements that are designed to provide more fluent authoring in some cases, but these enhancements are naturally less generic. Figure 1 illustrates this idea of LimSee3 seen as a generic authoring platform on which specific tools can be built. The development of specific tools is done thanks to a close interaction with a group of teacher users [4].

For instance, two specific tools have been developed during our collaboration with these users as a response to their needs:

- The *slideshow* creator allows users to build slideshows in a simple and efficient way. It is based on an dedicated document template which guides the user through the authoring process and provides some automation in the treatment. This specific tool was designed to respond to the need for easy preparation of a course material.

- The *multimedia course* builder is intended to be used after a course, to create a fully synchronized multimedia presentation out of the slideshow and the video and/or audio tracks shot during the lesson, with the possibility to provide additional annotations into the post-produced document. This tool addresses the need for production of on-line viewable course presentations.

Both tools were developed in collaboration with users, in a participatory-design way. User feedback validated our approach in that it proved that our two specific tools can be used as an authoring chain to prepare and reuse course material. However, a clear need for multiple presentation formats was also identified, the choice of a delivery format depending namely on the targeted audience. This need led us to develop a general mechanism for exporting into various presentation formats, presented in the following section.

## 4. ARCHITECTURE OF THE EXPORT SERVICE

The exportation process sketched below must cover the following objectives:
- Multiple targets, to cover user needs in publication and access formats.
- Optimization of the resulting document.
- Efficiency in the development of export modules for various formats.
- Extensibility, to cope with future formats.

This section is devoted to (1) the presentation of the exportation architecture and (2) the intermediate export format we have defined for that purpose.

### 4.1 Exportation Process

The exportation process can be illustrated with the case of exporting documents from LimSee3 documents to the SMIL format. As can be seen by comparing the document structure of SMIL with the LimSee3 document structure [7], the transformation requires to fully resolve and reorganize the time and spatial components (see Figure 2).

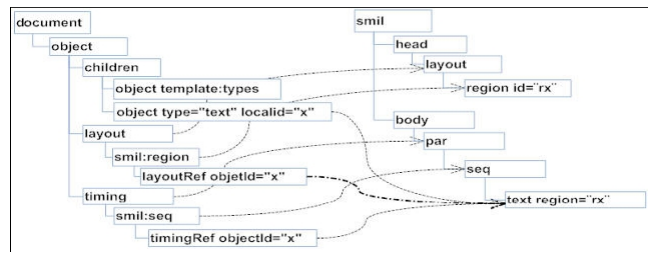

**Figure 2. Transformation of LimSee3 document structure into SMIL document structure**

Several solutions have been studied, from a pure XSLT-based one to a pure Java-oriented one. The first approach seems to be straightforward as both the LimSee3 syntax and the targeted publication formats are XML languages. However it has proven to be complex and inefficient because the XSLT code is not relevant for the required treatments such as time and spatial position computations. Moreover it does not allow to easily capitalize and share existing export services. Pure Java solutions benefit from the power of a programming language but imply that all new export development be done by a SAX/DOM developer. Finally, the proposed architecture of exportation takes advantage of both approaches: an intermediate format has been defined to convey all the structures and formatting parameters that can be computed by LimSee3 core modules; each targeted document format can be produced with an XSLT transformation (or a Java module) as shown in Figure 3. As will be discussed in conclusion, this architecture can be applied for other interchange needs which are not related to our LimSee3 authoring format. The important point to notice is that the intermediate format allows the capitalization of spatial and time computations of multimedia structures.

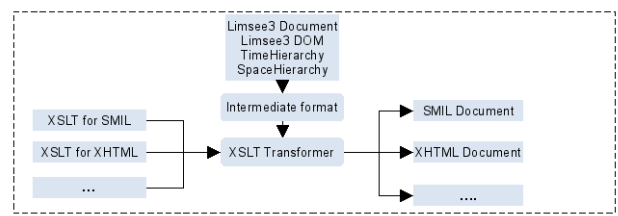

**Figure 3. The proposed export architecture using an intermediate format**

### 4.2 Intermediate Export Format

An intermediate format document is a valid XML document conforming to the following DTD:

http://ns.inria.fr/limsee3/intermediate/intermediate.dtd.

Elements describing the intermediate format are declared in the http://ns.inria.fr/limsee3/intermediate namespace. The root element is "document" which contains the following sections:

head
: It contains document meta-data.

layout
: This section describes a hierarchy of regions, organizing the actual presentational space of the document. To ensure target format agnosticism, each container corresponds to one displayable object – in particular there is no notion of region reuse as in SMIL.

timing
: It describes the time-container hierarchy much in the way the layout section does for space containers. The hierarchy is a tree formed by three different kinds of time containers: par (parallel-time container), seq (sequential-time container), excl (exclusive-execution container) and leaf container for actual media. The order of temporal objects in a sequence is important.

references
: This section is formed by a list of references. Each reference links an object (uniquely determined by its objectId) to a space container and a time container.

media
: This last section lists the basic media contained in the document; by linking abstract objects (represented by their objectId) referenced in other sections to actual media assets.

## 4.3 Intermediate Format Features

The overall objective of the intermediate format is to provide as much data as possible to subsequent transformation agents, while preserving all presentational semantics from the source document.

Therefore, an intermediate document would contain all statically computable information, in order to limit computational needs of subsequent agents. In fact, the intermediate format provides an unfolding (or a projection) of a source document on five different axis (meta-data, spatial layout, timing, internal dependencies, external dependencies).

The example in Figure 4 shows that for instance region information is utterly computed (all positioning attributes were resolved to pixel values). The timing tree (i.e. time container hierarchy) is produced, however some timing attributes cannot be statically known (they depend on the actual duration of the audio media) – players have to treat them dynamically.

We can notice that the intermediate format contains redundant or unnecessary information (such as precise time sequencing of a parallel time container). This is on purpose, since the transformation of a redundant document requires less special case handling, less value computation, less data-structure browsing than a non-redundant document.

This intermediate format aims at providing efficient export services for multimedia documents. Even if it does not yet cover a wide range of formats, it can be compared with formatting formats for static documents like XSL-FO that can be processed for the production of output formats such as ps, pdf, or rtf. Indeed, like XSL-FO, our intermediate format provides a way to store partial formatting information in a step-by-step formatting process and therefore provides a way to share this low-level information between several output processes. Instead of the format itself, our work aims at promoting such an approach for multimedia documents.

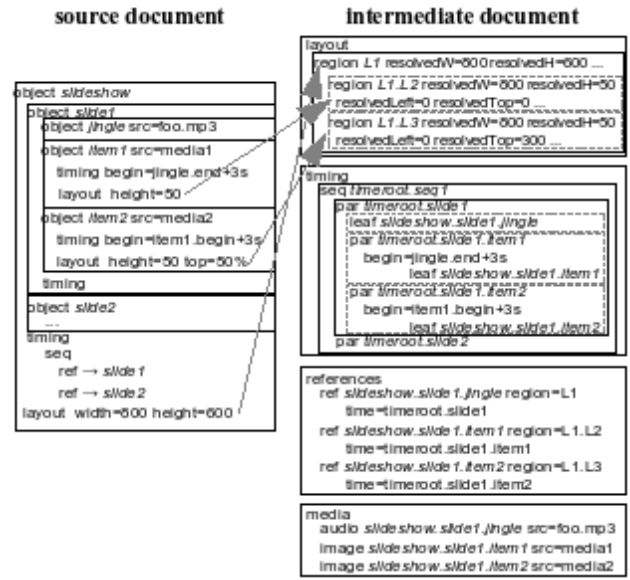

**Figure 4. Example of source and intermediate documents**

## 5. EXPERIENCING WITH THE EXPORT SERVICE

Implementing transformation from the LimSee3 document format into the intermediate format was straightforward, since every piece of information needed by an intermediate document is already present in the LimSee3 application, as part of some authoring service. For instance, the spatial layout hierarchy is used in LimSee3 to provide a static spatial view of the document, and as such it relies on resolved attributes values (coordinates, width, height,...). Thus, creating an intermediate document from a LimSee3 one is not much more than gathering known data and outputting it in an XML form.

We started creating transformations from the intermediary format to a presentation format with SMIL, which is the W3C standard for synchronized multimedia documents and which is probably the closest formalism to the LimSee3 document model (in terms of general approach to multimedia). In fact, transforming an intermediate document into SMIL proved to be easy: it consists in outputting the head, layout and timing sections in SMIL syntax, while omitting some data (e.g. the resolved spatial attributes, not needed by SMIL). One non trivial part is the on-the-fly resolution of references to actual media assets.

The intermediate-to-SMIL transformation procedure was implemented as a Java class and was extensively tested. Tests validated our approach in that the resulting SMIL documents are valid and are obtained in an efficient way. However, we were forced to adopt a modification when exporting some media objects, to ensure correct presentation behavior. It is a fact that the SMIL2 standard allows text objects, but does not rule on their formatting. Therefore, available players (RealOne, Ambulant Player) handle formatted text differently (e.g. RealOne defines a HTML-like syntactic extension to SMIL to allow text to be presented in a formatted way). This situation evolves with SMIL3, but no general-public player is available for that new standard yet. Since we intend to use SMIL as a presentation format only, we decided to export all text objects as PNG images when exporting to SMIL. This is currently the only way to ensure correct rendering semantics on all SMIL players. With this last amendment, our exporting approach becomes fully satisfactory.

The next targeted language is XHTML. While this language is not primarily designed for multimedia presentations, it is not forbidden either. We intend to benefit from the ubiquity of the web and from the constant evolution of web browsers (as compared to stagnant SMIL players). For documents requiring few synchronization features (such as slideshow presentations) this rendering format is clearly adapted. When more timing control is necessary, the use of some Javascript code has to be added. That is exactly what is proposed in the Timesheets specification [9] issued from SMIL3.0.

Here, the spatial structure of the intermediate document provides the main structure of XHTML (body) with absolute positioning (CSS). Time and interaction structures of the source documents are translated into timesheets elements. The XHTML player makes use of a Javascript scheduler for ensuring the correct behavior of the document as illustrated in Figure 5.

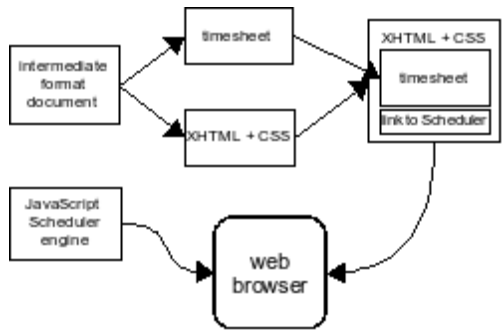

**Figure 5. Combining XHTML, CSS and Timesheets**

Such a Javascript engine has been proposed by P. Vuorimaa [10]. Its current version (0.5) implements the scheduling of static medias in a top-down manner: time containers handle displaying of their children. With some minor enhancements, we were able to experiment this XHTML+Timsheets+JavaScript engine approach on actual multimedia documents containing one continuous media and no user interaction.

Based on this work, we are currently implementing a more complete scheduler, including in particular the management of several continuous media (thanks to the VLC Mozilla plug-in) and taking into account various user interactions.

These production chains are being experienced by users to produce multimedia courses where the objective is to automate as far as possible content production and publication. One result is its use in the publication of a course in history on a publicly accessible course platform [11].

## 6. CONCLUSION

The benefits of the proposed intermediate format are twofold: it facilitates the deployment of authoring services independently from rendering systems and it simplifies the adoption of new technologies such as Timesheets because export features (basically transformation sheets) are easier to develop.

Moreover, as this intermediate format captures all the semantics of documents presentation, it can be considered as a pivot format between existing multimedia languages. The first application that can be rapidly developed is the rendering of a SMIL document inside a web browser: indeed, for the subset of SMIL relevant for its rendering inside a browser (i.e. ignoring features like prefetch), it is quite straightforward to produce the transformation from SMIL to our interchange format (except for the animation part that we have not yet implemented) and then apply the transformation towards XHTML with Timesheets.

A future step in the use of this architecture will be to consider more advanced publications needs such as those required for rendering adapted multimedia content, taking into account user needs or user context [1], [6].

## 7. ACKNOWLEDGMENTS

This work is supported by the Palette European project (FP6-028038).